\documentstyle[12pt]{article}
\input epsf
\def\boxit#1{\vbox{\hrule\hbox{\vrule{#1}\vrule}\hrule}}
\def\INSERTFIG#1#2#3{\epsfxsize=#1in \begin{center} \hbox
  to\hsize{\hfil\boxit{\epsffile{#2}}\hfil} {\sl #3} \end{center}}

\begin{document}
\begin{titlepage}
\begin{center}
January, 1995 \hfill CMU-HEP94-36\\
{}~  \hfill DOE/ER/40682-89\\
{}~  \hfill PITT-95-273\\
{}~  \hfill hep-ph/9501380\\
\vskip.7in {\Large\bf
Domain Walls in a FRW Universe}
\vskip.3in
D. Boyanovsky\footnote{Email: \tt boyan@vms.cis.pitt.edu}$^{(a)}$,
D. E. Brahm\footnote{Email: \tt brahm@fermi.phys.cmu.edu}$^{(b)}$,
A. Gonz\'alez-Ruiz\footnote{Email: {\tt alex@puhep1.Princeton.edu}; present
  address: Dept. of Physics, Princeton University, Princeton,
  New Jersey 08544}$^{(a)}$,\\
R. Holman\footnote{Email: \tt holman@fermi.phys.cmu.edu}$^{(b)}$, and
F. I. Takakura\footnote{Email: {\tt takakura@phyast.pitt.edu}; permanent
  address: Universidade Federal de Juiz de Fora, ICE, Depto.
  de F\'{\i}sica, Juiz de Fora, MG, Brazil.}$^{(a)}$\\~\\
{\it $^{(a)}$Dept.\ of Physics and Astronomy, University of Pittsburgh,
  Pittsburgh PA 15260}\\
{\it $^{(b)}$Carnegie Mellon Physics Dept., Pittsburgh PA 15213}

\end{center}

\vskip.2in
\begin{abstract}
We solve the equations of motion for a scalar field with domain wall boundary
conditions in a Friedmann-Robertson-Walker (FRW) spacetime.  We find (in
agreement with Basu and Vilenkin) that no domain wall solutions exist in de
Sitter spacetime for $h\equiv H/m \ge 1/2$, where $H$ is the Hubble parameter
and $m$ is the scalar mass.  In the general FRW case we develop a systematic
perturbative expansion in $h$ to arrive at an approximate solution to the field
equations.  We calculate the energy momentum tensor of the domain wall
configuration, and show that the energy density can become {\it negative\/} at
the core of the defect for some values of the non-minimal coupling parameter
$\xi$. We develop a translationally invariant theory for fluctuations of the
wall, obtain the effective Lagrangian for these fluctuations, and quantize them
using the Bunch-Davies vacuum in the de Sitter case.  Unlike previous analyses,
we find that the fluctuations act as zero-mass (as opposed to tachyonic) modes.
This allows us to calculate the distortion and the normal-normal correlators
for the surface.  The normal-normal correlator decreases logarithmically with
the distance between points for large times and distances, indicating that the
interface becomes rougher than in Minkowski spacetime.
\end{abstract}
\end{titlepage}
\setcounter{footnote}{0}


\section{Introduction}

The formation of topological defects in a cosmological phase transition is a
basic result in particle cosmology lore \cite{kibble,vilenkinrev}.  The
existence of defects provides us with a powerful tool which both constrains
particle theory models, such as those containing domain walls \cite{domainprob}
and magnetic monopoles \cite{monopoleprob}, and solves some outstanding
cosmological problems, such as the formation of structure with the use of
cosmic strings \cite{vilenkinrev,kibblemarsh}. Furthermore, during
out-of-equilibrium phase transitions, topological defects such as interfaces
(domain walls) are important ingredients for the dynamics of phase separation
and phase coexistence.

It has always been assumed that if a field theory defined in Minkowski
spacetime admits topological defects, the same will be true in an expanding
universe, and that furthermore, there will not be any significant differences
in the physical characteristics of these defects.  Recently, however, Basu and
Vilenkin \cite{basu} have shown that defects in a de Sitter background can have
properties that are quite different from defects in flat spacetime.  In this
work, we continue and extend this program by analyzing the equations of motion
for a domain wall in a background Friedmann-Robertson-Walker (FRW) spacetime.
We find that although these equations cannot be solved analytically in general,
a systematic perturbative expansion can be set up, with the small parameter $h
\equiv H/m$ being the ratio of the correlation length of the field to the
horizon size. Such a perturbative expansion arises naturally from the
requirement that the width of the domain wall be much smaller than the particle
horizon, to allow for kink boundary conditions within the microphysical
horizon.  The zeroth order solution is taken to be the standard flat spacetime
domain wall configuration, except that it is a function of the {\it physical\/}
spatial coordinate.  The higher order terms in $h$ are then solved for in a
systematic perturbative fashion.  We can also analyze the de Sitter case
numerically; we find that we agree with the results of Basu and Vilenkin
concerning the fact that for sufficiently large values of $h$, no wall
solutions exist -- when the horizon size is smaller than the domain wall width,
a kink configuration cannot fit in the horizon.

We next perform an expansion in the physical spatial coordinate that is
nonperturbative in $h$, and use it to examine the behavior of the energy
momentum tensor for the wall near the origin.  This is of interest due to
the possibility that a more natural form of inflation may occur in the core of
the defect, where the field is trapped in the unstable vacuum \cite{topinf}.

In curved spacetime, renormalization arguments suggest that, in
general, a coupling to the Ricci scalar has to be introduced in the bare
Lagrangian.  We find that such coupling is responsible for very remarkable
effects that result in a {\it change of sign} of the stress tensor at the
origin.  In fact, we see that this effect can also happen in flat spacetime, if
the so-called {\it improved\/} stress tensor \cite{jackiw} is used.

We derive the effective Lagrangian for the fluctuations perpendicular to the
wall. We find corrections beyond the ``Nambu-Goto'' action (the 3-volume swept
by the worldsheet of the wall) in a consistent expansion in derivatives.  We
also find non-trivial contributions to the surface tension from gravitational
effects.

We use this effective Lagrangian to address the question of whether the shape
of the wall fluctuates, or whether it remains flat at long distances and long
times after the formation of the defect.  Garriga and Vilenkin \cite{gavil}
have discussed this question to a certain extent, however we find that by using
collective coordinates the long-wavelength fluctuations are identified with
Goldstone modes and the instability found by them is avoided.

In the next section we set up our ansatz and perturbation expansion.  Section 3
deals with the energy momentum tensor for the wall in an FRW universe.  We find
that interesting behavior can occur by making use of the coupling $\xi {\cal R}
\phi^2$ of the field to the Ricci scalar ${\cal R}$.  In section 4, we analyze
fluctuations of the wall.  We differ from Garriga and Vilenkin in our treatment
of translational invariance.  The fact that the wall is located at a particular
point along the $z$-axis, say, would appear to break this symmetry.  However,
translational invariance can be restored by use of the collective coordinate
method.  Doing this, we arrive at a true Nambu-Goldstone mode for the
fluctuations, as opposed to one with negative ${\rm mass}^2$ as found by
Garriga and Vilenkin.  Several equal (comoving) time correlation functions that
determine the behavior of the shape of the wall are computed, and we find that
their spatial fall-off is much slower than in flat Minkowski spacetime at long
times.  This leads us to the conclusion that the surface remains ``rough'' at
large separations.

Section 5 contains our concluding remarks and some cosmological implications of
our results.

\section{The Domain Wall}

Domain walls (interfaces) are field configurations with non-trivial
topological boundary conditions.  These boundary conditions demand that the
field vary substantially within a typical spatial scale that is usually
determined by the (finite temperature) correlation length $\xi \approx m^{-1}$,
with $m$ the, in general temperature-dependent, effective mass of the field.

In FRW cosmologies there is another important scale, the microphysical horizon
size $d_h = H^{-1}$ where $H$ is the Hubble parameter.  Causality implies that
the region in which the scalar field can vary appreciably must be subhorizon
sized, since only then can the boundary conditions that define the wall fit
inside the horizon.  Thus we expect that the notion of a domain wall will only
make sense if $H/m \leq 1$.

When $H/m \ll 1$, we expect that the domain wall profile will exhibit only
small deviations from the Minkowski spacetime profile.  We can then study the
differences in a power series expansion in $H/m = \xi / d_h $.

Consider a scalar field $\phi$ with potential
\begin{equation}
 U = {\lambda\over4} (\phi^2 - v^2)^2 + {\xi\over2} {\cal R} \phi^2
\end{equation}
in a FRW spacetime with metric $g_{\mu\nu}={\rm diag}(1,-a^2,-a^2,-a^2)$, and
Hubble parameter $H \equiv \dot{a}\slash a$ (where $\dot{a} \equiv {\rm d}a/
{\rm d}T$).  The second term couples the field to the Ricci scalar ${\cal R} =
6(2 H^2 + \dot{H})$.  Such a term is required for consistency as it will be
generated by quantum corrections in general; the only fixed point is $\xi =1/6$
which corresponds to conformal coupling \cite{boyveghol}.  The equation of
motion is
\begin{equation}
 {\partial^2\phi\over\partial T^2} + 3 H {\partial\phi\over\partial T} -
 {1\over a^2} \nabla^2 \phi + {\partial U\over\partial\phi} = 0
\label{eomc} \end{equation}
where $\nabla$ is the derivative with respect to {\it co-moving\/} spatial
coordinates $\{X,Y,Z \}$.  It is convenient to define the following
dimensionless quantities
\begin{eqnarray}
 && m = \sqrt\lambda\,v, \qquad \{x,y,z,t\} = m \{X, Y, Z, T \}, \qquad h =
  H/m, \nonumber\\
 && \eta = \phi/v, \qquad \omega=a(t)z, \qquad \beta=4+\dot{H}/H^2, \qquad
 \mu^2 = 1 - 6\xi h^2 (\beta-2).
\label{dimenquan} \end{eqnarray}

Note that $\omega$ is a (dimensionless) physical coordinate.  For power-law
expansion (PLE) $a(t) = t^n$, we have $\beta = 4\!-\!1\slash n$.  In the
radiation dominated (RD) and matter dominated (MD) flat universes respectively,
$\beta_{RD} = 2$ and $\beta_{MD} = 2.5$.  In de Sitter spacetime $\beta_{dS}=4$
(or equivalently, we can take $n=\infty$).

We will assume a domain wall (kink) along the $z$-axis and that the
corresponding field configuration is independent of the transverse coordinates,
so that we are considering a flat interface.

The scale factor dividing the (comoving) Laplacian in eq.~(\ref{eomc}) suggests
the following ansatz for the kink configuration:
\begin{equation}
 \eta(z,t)= \eta(\omega,h(t)).
\end{equation}

Such a solution obeys
\begin{equation}
 (1 - h^2 \omega^2) \eta'' - \beta h^2 \omega \eta' + \mu^2 \eta - \eta^3 =
 \frac{1}{n^2} h^4 {\partial^2 \eta\over\partial h^2} - \frac{2}{n} h^3 \omega
 {\partial \eta'\over \partial h} + \left(\frac{2}{n^2} - \frac{3}{n}\right)
 h^3 {\partial \eta\over\partial h}
\label{static} \end{equation}
where a prime means $\partial/\partial\omega$.  The right-hand side (RHS)
vanishes for de Sitter spacetime, and is ${\cal O}(h^4)$ for PLE.

In the case of de Sitter spacetime, the effect of non-zero conformal coupling
$\xi$ ({\it i.e.\/} $\mu\ne 1$) can be absorbed in a simple rescaling of
variables:
\begin{equation}
 \eta=\mu\bar\eta, \quad \omega=\bar\omega/\mu, \quad h=\mu\bar h.
\label{scale} \end{equation}
For PLE, this introduces additional ${\cal O}(h^4)$ terms on the RHS of
eq.~(\ref{static}), since $\dot{\mu} \ne 0$.

Except to discuss the energy-momentum tensor $T_{\mu\nu}$, we will hereafter
work in the rescaled theory (and drop the bars), writing
\begin{equation}
 (1 - h^2 \omega^2) \eta'' - \beta h^2 \omega \eta' + \eta - \eta^3 = \left\{
 \matrix{0 \hbox{~(de Sitter)} \cr {\cal O}(h^4) \hbox{~(PLE)} \cr} \right.
\label{stat2} \end{equation}
and taking ``soliton boundary conditions'' $\eta(\pm\infty, h) = \pm 1$.  All
results for PLE are thus valid only to ${\cal O}(h^2)$.

\subsection{Walls in de Sitter Spacetime: Numerical Solution}

The differential equation (\ref{static}) is, in general, a partial differential
equation in two variables, and not easily amenable to numerical study.  As we
see in eq.~(\ref{stat2}), ignoring the RHS would introduce errors at ${\cal
O}(h^4)$ for PLE.  However, in de Sitter spacetime, where $h$ is constant, we
have an ordinary differential equation, for which we can find exact numerical
solutions.  Indeed, this was done by Basu and Vilenkin \cite{basu}, and our
results appear to agree with theirs.

Our results are plotted as thin curves in Fig.~1.  The slope at the origin was
chosen (by a shooting algorithm) by requiring smoothness through the point
$\omega=1/h$ (marked with a tick).  As found by Basu and Vilenkin \cite{basu},
solutions only exist when $h^2 < 1/4$.  (For PLE, if one insists on ignoring
the RHS of eq.~(\ref{stat2}), solutions only exist when $h^2 < 1/\beta$.)

We see the kink has been flattened out by the effects of the cosmological
expansion.  The slope at the origin $a_1 \equiv \eta'(0,h)$ decreases with
$h^2$ as shown in Fig.~2, vanishing at $h^2 = 1/4$.  In other words, for a
topologically stable solution to exist, the horizon size must be greater than
the correlation length (by a factor of 2).

\subsection{Expansion in $h^2$}

Although a numerical solution is available in the case of de Sitter, the
general case is extremely difficult to analyze numerically, since we now have a
non-linear partial differential equation in two variables.  However, when $h^2
\ll 1/\beta$, we can solve eq.~(\ref{stat2}) in a systematic perturbative
expansion in $h^2$, and the first-order result will be valid for both de Sitter
and PLE cosmologies.  We write
\begin{equation}
 \eta(\omega,h) = \eta_s(\omega) + \sum_n h^{2n} \delta^{(n)}(\omega) \;.
\end{equation}

Here $\eta_s$ is the kink configuration which solves eq.~(\ref{stat2}) for
$h=0$ (flat spacetime):
\begin{equation}
 \eta_s(\omega) = \tanh(\omega/\sqrt2)
\label{kin} \end{equation}
This $h=0$ kink is shown as a dashed curve in Fig.~1.  Note that $\eta'_s(0) =
1/\sqrt{2}$.

Since $\eta_s$ has the correct asymptotic behavior, we must set
$\delta^{(n)}(\pm\infty)=0$.  To leading order in $h^2$, $\delta(\omega)$ obeys
\begin{equation}
\hat{O}_s\delta^{(1)} =
 \frac{d \delta^{(1)}}{d \omega^2} + \delta^{(1)} - 3 \eta_s^2(\omega)
 \delta^{(1)} = j(\omega) \equiv
 \eta_s''(\omega) \omega^2 + \beta \eta_s'(\omega) \omega \;. \label{fluct}
\end{equation}
The solution may be found by elementary methods once a solution of the
homogeneous equation is found.  Such a solution is easily available, since
translational invariance of the unperturbed differential equation
guarantees that
\begin{equation}
 \delta_1(\omega) = \eta'_s(\omega)=\frac{1}{\sqrt{2}} {\rm
 sech}^2\left(\frac{\omega}{\sqrt{2}} \right),
\end{equation}
is a ``zero mode'' of the operator $\hat{O}_s$ of quadratic fluctuations
around the kink configuration.

Another linearly independent solution $\delta_2(\omega)$ having unit Wronskian
with $\delta_1(\omega)$ can be found by elementary methods:
\begin{equation}
 \delta_2(\omega) \equiv {1\over4} \left[\sinh\left( \sqrt2 \omega \right) + 3
 \tanh\left( \omega\over\sqrt2 \right) + {3\omega\over\sqrt2} {\rm
 sech}^2\left(\omega\over\sqrt2 \right) \right],
\end{equation}
where we have chosen this solution to vanish at the origin.
Finally the first order solution  can be written
\begin{equation}
 \delta^{(1)} (\omega) = \delta_2(\omega) \int_0^\omega \delta_1(\zeta)
 j(\zeta) {\rm  d}\zeta - \delta_1(\omega) \int_0^\omega \delta_2(\zeta)
 j(\zeta) {\rm d}\zeta
 +a \delta_1(\omega)+b \delta_2(\omega), \label{fullsol}
\end{equation}
where the coefficients $a,b$ have to be determined from the boundary
conditions.  Notice that $\delta_1(\omega)$ is symmetric whereas $j(\omega), \;
\delta_2(\omega)$ are antisymmetric functions of $\omega$.

Note that $\delta_2(\omega)$ diverges exponentially as $\omega \rightarrow
\pm \infty$.  To render the solution finite and to satisfy the boundary
conditions, we must choose:
\begin{equation}
 b = -\int^\infty_0 \delta_1(\zeta)j(\zeta) d\zeta . \label{bcoeff}
\end{equation}

The boundary condition at infinity {\it does not} determine the constant
$a$. Notice that the term $a \delta_1(\omega)$ represents a local
translation of the wall.  Clearly the freedom of choice of $a$ reflects
the underlying translational invariance.
If we demand that
the wall position be at the origin,  we must demand that
$\delta^{(1)} (\omega=0)=0$; this leads to $a=0$.

Thus the final solution to this order that satisfies the boundary conditions
and keeps the wall centered at the origin is given by
\begin{equation}
 \delta^{(1)} (\omega) = -\delta_2(\omega) \int_\omega^\infty \delta_1(\zeta)
 j(\zeta)
 {\rm d}\zeta - \delta_1(\omega) \int_0^\omega \delta_2(\zeta) j(\zeta) {\rm
 d}\zeta . \label{finalwall}
\end{equation}

The results for de Sitter spacetime ($\beta=4$) are shown as thick curves in
Fig.~1.  $\delta^{(1)}(\omega)$ for $\beta=4$ is plotted in Fig.~3.  This
approximation underestimates the distortion away from the $h=0$ kink, and does
not see the singularity at $h^2=1/4$, which is well beyond the regime of
validity of the perturbative expansion in $h^2$.

It is straightforward to see that the structure of the perturbative expansion
persists to all orders.  In fact we find that
\begin{equation}
 \hat{O}_s \delta^{(n)} = j^{(n)}
\end{equation}
with the same differential operator $\hat{O}_s$ as in (\ref{fluct}), and where
the source term $j^{(n)}$ is obtained from the solutions up to order
$n-1$. Clearly the same functions $\delta_1(\omega) \,,\, \delta_2(\omega)$
generate the solutions, as these are the two linearly independent solutions of
the homogeneous differential equation $\hat{O}_s \delta = 0$. Thus the general
form of the solution is the same as (\ref{fullsol}) with $j^{(n)}$ replacing
$j$ and coefficients $a^{(n)},b^{(n)}$ replacing $a,b$.

By the same arguments presented before, a finite solution at infinity requires
that
\begin{equation}
 b^{(n)} = -\int^\infty_0 \delta_1(\zeta)j^{(n)}(\zeta) d\zeta
 . \label{bncoeff}
\end{equation}

The source term $j^{(n)}(\omega)$ is constructed from iterations up to order
$(n-1)$, and for $n=1$ it is given by derivatives of the unperturbed kink
solution that vanish at $\omega = \pm \infty$. It is straightforward to prove
by induction that the source term $j^{(n)}(\omega)$ is antisymmetric and
vanishes exponentially as $\omega \rightarrow \pm \infty$. Therefore $b^{(n)}$
is finite to all orders.  Translational invariance suggests that $a^{(n)}=0$ as
before, and the the solution to order $n$ is antisymmetric and this property
ensures that the source term for the next order iteration is antisymmetric
again.  Thus we are led to the conclusion that the general form of the
perturbative solution is given by eq.~(\ref{finalwall}) with $j(\omega)$
replaced by $j^{(n)}(\omega)$, and the only complication in carrying out this
program to any arbitrary order is to find the source term iteratively.  Clearly
this will be a highly non-local function because of the nested
integrals, and the numerical evaluation will become more cumbersome in higher
orders, but in principle this scheme will yield a consistent perturbative
expansion.  Clearly we have no way of determining the radius of convergence of
such an expansion.

\subsection{Expansion in $\omega$}

The field configuration corresponding to a domain wall has most of the gradient
and potential energy difference (with respect to the broken symmetry vacuum)
localized in a spatial region of the order of the correlation length.  Most of
the contribution to the energy momentum tensor, after subtracting off the
vacuum value, will arise from this small region around the position of the
domain wall.  This motivates us to obtain the kink profile near the origin in a
power-series expansion in the coordinate $\omega$.  This expansion is {\it
nonperturbative\/} in $h^2$.

The motivation for this expansion is the following: one would eventually like
to study the full system of Einstein's equations and matter field in a
self-consistent semiclassical manner.  In particular we have in mind the
question of how the presence of domain walls affect the gravitational
fields.  Since for this task one only needs the behavior of the energy momentum
tensor, and as we argued above most of the contribution to its components from
a domain wall field configuration arise from a very localized region near the
position of the interface, a power series expansion near the origin may capture
most of the relevant physical features for an understanding of the back
reaction of domain walls on the gravitational field.  This argument becomes
more relevant in view of the proposal of rapid inflation at the core of
topological defects \cite{topinf}.

In the general case of de Sitter or power-law expansion we find the following
behavior of the interface profile near the origin:
\begin{equation}
 \eta(\omega,h) = a_1 \left[ \omega + {\beta h^2 - 1 \over 6} \omega^3 + \left(
 {(\beta h^2 - 1) ([6+3\beta] h^2 - 1) \over 120} + {a_1^2 \over 20} \right)
 \omega^5 + {\cal O}(\omega^7) \right]
\label{orig} \end{equation}
The linear coefficient $a_1$ cannot be determined perturbatively, and
will have to be found by solving the differential equation, its value being
determined by requiring smoothness through $\omega = 1/h$.  We notice,
however,that as $h^2\to 1/\beta$ from below, the first term dominates (even for
large $\omega$) and the profile flattens out to a straight line.

Since this expansion works best for $h^2$ close to $1/\beta$, but is correct
only to ${\cal O}(h^2)$ for PLE, it is only useful in this case very close to
the origin.

Given a value for $a_1$ (from the numerical solution) as input, we display this
series [through ${\cal O}(\omega^5)$] for the de Sitter spacetime ($\beta=4$)
case as the thin dot-dashed curve in Fig.~1.  For $h^2 \mathrel{\scriptstyle{
\buildrel > \over \sim}} 0.2$, where the solution is very flat, the
small-$\omega$ expansion is valid out to well past $\omega=1/h$. We will
use these results in the following section.

\section{The Energy-Momentum Tensor}

We next turn to an examination of the energy momentum tensor for the domain
wall profile in an FRW spacetime.

In general, the energy-momentum tensor of a scalar field configuration is given
by
\begin{equation}
 T_{\mu\nu} = {2 \over\sqrt{-g}} {\delta (\sqrt{-g} {\cal L})\over \delta
 g^{\mu\nu}}, \qquad {\cal L} = {1\over2} g^{\alpha\beta} (\partial_\alpha
 \phi) (\partial_\beta \phi) - U(\phi)
\label{Tmunu} \end{equation}
where $g \equiv \det\{ g_{\mu\nu} \}$, and we use
\begin{equation}
 {\delta \sqrt{-g} \over \delta g^{\mu\nu}} = {-\frac{1}{2} \sqrt{-g}}
 \, g_{\mu\nu}
\end{equation}

At this stage we restore the non-minimal coupling to the Ricci scalar, since,
as we argued above, such a term will be induced by renormalization if not
present in the original Lagrangian \cite{boyveghol}.  Its presence has two
major effects: the first one is to modify the form of the energy momentum
tensor, while the second is to modify the field configuration for the domain
wall.  Unlike the analysis of the equation of motion in the previous sections,
for the case of the energy momentum tensor we can no longer rescale variables
to get rid of $\xi$, since $(\delta {\cal R} / \delta g^{\mu\nu}) \ne 0$.
Let $U_0$ be the potential for $\xi\!=\!0$, so $U=U_0 + (\xi/2) {\cal R}
\phi^2$.  Performing the variation of eq.~(\ref{Tmunu}), we find:
\begin{eqnarray}
 T_{\mu\nu} &=& (1-2\xi) \phi_{;\mu} \phi_{;\nu} + (2\xi-{\textstyle{1\over2}})
  g_{\mu\nu} g^{\alpha\beta} \phi_{;\alpha} \phi_{;\beta} + g_{\mu\nu} U_0
  \nonumber\\
 &&\qquad - \xi \phi^2 (R_{\mu\nu} - {\textstyle{1\over2}} g_{\mu\nu}R) + 2
  \xi \phi (g_{\mu\nu} \Box - \nabla_\mu \nabla_\nu) \phi
\end{eqnarray}
In our FRW metric, the non-zero components of $t_\mu^{\;\nu} \equiv
T_\mu^{\;\nu}/(m^2 v^2)$ are:
\begin{eqnarray}
t_0^{\;0} &=& {1\over2} \left[ 1+(h\omega)^2 \right] (\eta')^2 + {1\over4}
  (\eta^2-1)^2 \nonumber\\
 &+& \xi \left\{ 3 h^2 \eta^2 + 6 h^2 \omega \eta\eta' - 2
   \left[(\eta')^2 + \eta\eta''\right] \right\} \nonumber\\
t_1^{\;1} = t_2^{\;2} &=& {1\over2} \left[ 1-(h\omega)^2 \right]
  (\eta')^2 + {1\over4} (\eta^2-1)^2 \nonumber\\
 &+& \xi \left\{ (2\beta-5) h^2 \eta^2 + (2\beta-2) h^2 \omega \eta
  \eta' - 2 \left[ 1-(h\omega)^2 \right] \left[(\eta')^2 + \eta\eta''\right]
  \right\} \nonumber\\
t_3^{\;3} &=& {-1\over2} \left[ 1+(h\omega)^2 \right] (\eta')^2
  + {1\over4} (\eta^2-1)^2 \nonumber\\
 &+& \xi \left\{ (2\beta-5) h^2 \eta^2 + (2\beta-2) h^2 \omega
  \eta\eta' + 2 (h\omega)^2 \left[(\eta')^2 + \eta\eta''\right] \right\}
  \nonumber\\
t_3^{\;0} = -a^2(t)\, t_0^{\;3} &=& a(t) h\omega \left\{ (\eta')^2 - 2 \xi
  \left[ (\eta')^2 + \eta \eta'' \right] \right\}
\label{emt} \end{eqnarray}
As usual, for PLE there are corrections at ${\cal O}(h^4)$.

The presence of the non-linear coupling to the Ricci scalar modifies the
definition of the energy-momentum tensor.  Even in the case of Minkowski
spacetime, with $h\!=\!0$, this modification of the energy-momentum tensor
produces some peculiar behavior in $t_\mu^{\;\nu}$ near the origin, despite the
fact that this coupling $\xi$ does not affect the equations of motion.  In this
case, the wall solution is just $\eta_s$ from eq.~(\ref{kin}), and the non-zero
components of $t_\mu^{\;\nu}$ are:
\begin{equation}
 t_0^{\;0} = t_1^{\;1} = t_2^{\;2} = \left[ \frac12 - 2\xi + \xi \cosh(\sqrt2
 \omega) \right] {\rm sech}^4(\omega/\sqrt2)
\label{t00flat} \end{equation}
For large enough $\xi$, the energy density is negative at the origin:
$(t_0^{\;0})(0) \le 0$ for $\xi \ge 1/2$.  For $\xi<0$, a region of negative
energy density occurs away from the origin.

We can substitute the small-$\omega$ expansion $\eta = a_1 \omega + {\cal
O}(\omega^3)$ (which is correct even for $\xi\ne 0$) into eq.~(\ref{emt}) to
find the behavior at the origin.  The non-zero components of $t_\mu^{\;\nu}(0)$
are:
\begin{equation}
 t_0^{\;0}(0) = t_1^{\;1}(0) = t_2^{\;2}(0) = \frac14 + \frac{a_1^2}{2} - 2
  a_1^2 \xi,  \qquad t_3^{\;3}(0) = \frac14 - \frac{a_1^2}{2},
\end{equation}
and the energy density at the origin is negative for
\begin{equation}
 \xi > \frac14 + \frac{1}{8 a_1^2} \;.
\end{equation}
These results are consistent with eq.~(\ref{t00flat}),
where $a_1^2 = \frac12$ [see equation (\ref{kin})].

In Fig.~4 we show exact numerical calculations of $t_0^{\;0}$ for de Sitter
spacetime with $h^2 \ne 0$.  The behavior of $t_0^{\;0}$ for small values of
$h^2$ (such as $h^2=0.02$ in Fig.~4a) is qualitatively similar to
eq.~(\ref{t00flat}).  For larger $h^2$ (such as $h^2=0.10$ in Fig.~4b) another
effect enters, namely that the wall profile $\bar\eta(\bar\omega)$ is really
determined by $\bar h = h/\mu$, as defined in eq.~(\ref{scale}).  Wall
solutions in de Sitter spacetime only exist for $\bar h^2 < 1/4$, which implies
\begin{equation}
 h^2 < {1 \over 4 + 12\xi}
\end{equation}
For example, with $h^2=0.10$, the wall profile flattens away to $\eta=0$ as
$\xi\to 1/2$ (where $\bar h^2 \to 1/4$), and $t_0^{\;0}(\omega) = 1/4$.

The lesson that we learn from this analysis is that the coupling to the Ricci
scalar can dramatically modify the behavior of the energy momentum tensor near
the origin, and may be an interesting possibility for topological inflation at
the core of defects \cite{topinf}.  These effects can even arise using the
``improved'' energy momentum tensor \cite{jackiw} in Minkowski spacetime, if
one starts with a general curved spacetime and then takes the flat limit.

\section{Fluctuations of the Wall}

Up to this point our study has focused on the description of a ``flat''
interface or domain wall, that is the field profile varies only along the
direction perpendicular to the domain wall but is constant on the perpendicular
directions. However there will be fluctuations both quantum and thermal that
will tend to distort locally the interface. An important question to address is
the following: are these fluctuations ``small'' in the sense that the wall
remains flat at long distances, or are the fluctuations important so that the
wall becomes ``rough''?  We will assume that the system is at zero temperature
and that only quantum fluctuations are important.

We begin by deriving the effective action for the fluctuations of the
interface and then proceed to calculate relevant correlation functions.

We depart from the treatment of Garriga and Vilenkin \cite{gavil}, in that we
sacrifice {\it explicit\/} covariance to treat translational invariance in
terms of collective coordinates; however, the final result for the action will
be fully covariant.  This procedure, borrowed from the usual scheme to quantize
the collective coordinates associated with translations of soliton
solutions \cite{raja} has many advantages.

The position of the interface explicitly breaks translational invariance.
However, a rigid translation of the interface should cost no energy due to the
underlying translational symmetry.  Thus the fluctuations perpendicular to the
interface should be represented by {\it massless\/} degrees of freedom since,
locally, they represent translations of the interface.  These are the capillary
waves or Goldstone bosons \cite{jasnow,safran,wallace} associated with the
breakdown of the translation symmetry.  Translational symmetry is then restored
by quantizing these fluctuations as collective coordinates \cite{raja}.

In terms of dimensionless comoving coordinates and the dimensionless field
$\eta$, the action is (${\rm d}^4 x$ is the comoving volume element in
dimensionless units)
\begin{equation}
 I=\frac{1}{\lambda}\int {\rm d}^4 x\; a^{3}(t)
 \left\{\frac{1}{2} \left(\frac{\partial \eta}{\partial t}\right)^2-
 \frac{1}{2a^2(t)}\left(\nabla \eta \right)^{2}
 - \frac{1}{4}\left(\eta^2-1 \right)^2 \right\}.
\end{equation}

We want to incorporate fluctuations of the interface solution (kink) and to
obtain the effective action for the long-wavelength modes.  In order to
understand how to achieve this goal, it proves convenient to recall how this
procedure works in Minkowski spacetime.

\subsection{Goldstone Modes in Flat Spacetime}

Consider first a static kink along the $z$-direction.  As is well
known \cite{raja} there is a zero mode of the linear fluctuation operator
\begin{equation}
 \delta_0\eta \approx \frac{\partial \eta_s(z-z_0)}{\partial z}
\label{zeromode} \end{equation}
where $z_0$ is the position of the kink and $\eta_s$ is the kink solution.
This mode corresponds to a translation of the position of the kink, because
$\eta_s(z-z_0) +\alpha\, \delta_0 \eta \approx \eta_s(z-z_0+\alpha)$.
Translational invariance guarantees that such a perturbation does not change
the energy and thus is a zero mode of the linear fluctuation operator.  Now
consider a kink in three space dimensions, corresponding to a wall along the
$z$-axis and centered at $z_0$.  A perturbation of the form
\begin{equation}
 \delta \eta_p(x,y,z-z_0) = a_p \frac{\partial \eta_s(z-z_0)}{\partial
 z}e^{i\vec{p}_\perp\cdot \vec{x}_\perp} \;,
\label{ppert} \end{equation}
with $\vec{p}_\perp\cdot \vec{x}_\perp = p_xx+p_yy$, is an eigenmode of the
linear fluctuation operator with eigenvalue $\omega_p=\sqrt{p_x^2+p_y^2}$
\cite{jasnow}.  These fluctuations of the interface correspond to the Goldstone
bosons of the broken translational symmetry, and are the capillary waves of the
interface \cite{jasnow,safran,wallace}.  The perturbed solution
\begin{eqnarray}
 & & \eta_s(z-z_0)+\sum_p \delta \eta_p(x,y,z-z_0) \approx
 \eta_s(z-f(x,y))\label{wallpert} \\
 & & f(x,y) = z_0-\sum_p a_p e^{i\vec{p}_\perp\cdot \vec{x}_\perp} \nonumber
\end{eqnarray}
corresponds to a {\it local translation\/} of the interface.  The $a_p$
correspond to ``flat directions'' in function space.  Just as in the
one-dimensional case, these Goldstone modes cannot be treated in perturbation
theory, because arbitrarily large $a_p$ for $\vec{p}_\perp \rightarrow 0$ can
be accessed at no cost in energy.  We will borrow results from the
one-dimensional procedure and quantize these modes as ``collective
coordinates'' \cite{sakita}.

Besides these Goldstone modes there are massive modes corresponding to the
higher energy states of the one-dimensional kink with dispersion relation
$E(p_\perp)=\sqrt{p_\perp^2+3m^2}$ \cite{raja,jasnow}.  Because of this gap in
the energy spectrum, we can safely concentrate on the long-wavelength
fluctuations of the interface and obtain an effective action for these
Goldstone modes, treated as collective coordinates.  The coordinates $a_p$ or
the field $f(x,y)$ are now fully quantized.  In a path integral quantization
procedure, collective coordinate quantization amounts to a functional integral
over all configurations of $f(x,y)$.  Clearly this procedure restores
translational invariance because now the field is invariant under $z
\rightarrow z+\delta, \ f(x,y) \rightarrow f(x,y)+ \delta$ and the field
$f(x,y)$ is now functionally integrated with a translational invariant measure.
Passing from the original integration variables in the functional integral to
the new variables including $f(x,y)$ involves a Jacobian which is seen to be
unity to the order that we are working in (see below).  For a thorough
exposition of collective coordinate quantization in the path integral and
Hamiltonian forms, the reader is referred to the original literature
\cite{raja,sakita,chrislee,tomboulis}.

\subsection{Goldstone Modes in a FRW Cosmology}

After this digression in Minkowski spacetime we are ready to extend these
observations to a FRW cosmology.  Let $\eta_s(\omega)$ be a solution to
eq.~(\ref{stat2}) (the $h$-dependence will now be implicit), and let us take
the small-gradient limit, in which
\begin{equation}
 f_{xx},\; f_{yy},\; f_{tt} \ll 1 \; ; \qquad hf_x,\; hf_y,\; hf_t \ll 1
\end{equation}
($f_{xx} \cdots$ are second derivatives with respect to the dimensionless
comoving coordinate $x$, {\it etc}.)  This approximation is consistent with our
purpose of studying the long-wavelength fluctuations on distances such that $1
\ll x,y \ll h^{-1}$.  We look for a profile of the form
\begin{equation}
 \eta(z,f(x,y,t),t) = \eta_s(\xi(z,x,y,t))
\label{etachi} \end{equation}
Note that the substitution $\xi(z,x,y,t) = a(t)[z-f(x,y,t)]$ does {\it not\/}
in general solve the equations of motion, but
\begin{equation}
 \xi(z,x,y,t) = a(t) \frac{\left[z-f(x,y,t)\right]}
 {\sqrt{1+f_{x}^{2}+f_{y}^{2}- a^{2}(t)f_{t}^{2}}}
\label{goch} \end{equation}
does, in our small-gradient approximation.

The denominator in eq.~(\ref{goch}) has an important physical meaning.  The
function $f(x,y,t)$ determines the position of the domain wall (interface).
This function induces a metric $g^{(3)}_{ab}$ on the $2+1$ dimensional
world-volume swept out by the wall, and we find, in terms of dimensionless
comoving variables,
\begin{equation}
 \sqrt{g^{(3)}}={a^2(t)}\sqrt{1+f_{x}^{2}+f_{y}^{2}-
 a^{2}(t)f_{t}^{2}}.
\label{inducedmetric} \end{equation}

The effective action for the displacement field $f(x,y,t)$ in the
long-wavelength approximation, is found by
following the usual procedure \cite{jasnow,safran,wallace} which consists of
computing the action for the profile (\ref{etachi}) with (\ref{goch}) (this is
also identified with the effective action for the collective
coordinate \cite{sakita}).  After integrating by parts and discarding
surface terms, we find
\begin{eqnarray}
 I&=&-\frac{1}{\lambda}\int {\rm d}^3 x\; a^{2}(t)\left[C_{0}-C_{2}
  (5+2\frac{\dot{h}}{h^{2}})\frac{h^{2}}{2}\right]
  \sqrt{1+f_{x}^{2}+f_{y}^{2}-a^{2}(t)f_{t}^{2}} \nonumber\\
 &+&\frac{C_{2}}{8\lambda}\int  {\rm d}^3 x\;
  \frac{a(t)^{2}}{\left(1+f_{x}^{2}+f_{y}^{2}-a^{2}(t)f_{t}^{2}\right)^{3/2}}
  \label{act} \\
 &&\left\{\left[\partial_{t}\left(1+f_{x}^{2}+f_{y}^{2}-a^{2}(t)f_{t}^{2}
  \right) \right]^{2}-\right.\nonumber\\
 &&\left.\frac{\left[\partial_{x}\left(1+f_{x}^{2}+f_{y}^{2}-a^{2}(t)f_{t}^{2}
  \right)\right]^{2}}{a(t)^{2}}-\frac{\left[\partial_{y}\left(1+f_{x}^{2}+
  f_{y}^{2}-a^{2}(t)f_{t}^{2}\right)\right]^{2}}{a(t)^{2}}\right\}, \nonumber
\end{eqnarray}
where ${\rm d}^3 x$ represents the $2+1$ dimensional comoving volume element,
and we have used $\eta_s(-\xi)=-\eta_s(\xi)$ to eliminate the integrations
linear in $\xi$ as well as having defined:
\begin{eqnarray}
 C_{0} &=& \int d\xi\left[\frac{1}{2}\left(\frac{\partial \eta_s}{\partial
  \xi} \right)^2+\frac{1}{4}(\eta_s^2-1)^2\right]\nonumber\\
 C_{2} &=& \int d\xi\left(\frac{\partial\eta_s}{\partial \xi}\right)^2 \xi^2
  .\nonumber
\end{eqnarray}

This is the final form of the action for the fluctuations of the interface.  It
exhibits the translational symmetry explicitly, and only contains derivative
terms as is required of a Goldstone field.
$C_0$ is identified with the (usual) flat spacetime surface tension, and
we see that curved spacetime effects induce a renormalization of this
surface tension. The first term (proportional to $\sqrt{g^{(3)}}$) is
recognized as the equivalent to the ``Nambu-Goto'' action, which is
essentially the total ``world-volume'' associated with the fluctuation
field. The second term is thus a correction to the ``Nambu-Goto'' action;
further corrections can be obtained in a systematic expansion in
derivatives for the solution of the equations of motion.

Expanding $I$ to quadratic order in
the fluctuation field shows that the fluctuations are {\it massless\/}:
\begin{equation}
 I_{\rm quad} = I_0 + \frac{1}{\lambda} \int {\rm d}^3 x\; a^4(t) \left[C_0 -
 C_2 (5+2\dot{h}/h^2) h^2/2 \right] \left[ \frac12 f_t^2 - \frac{1}{2a(t)^2}
 (f_x^2 + f_y^2) \right] \;,
\label{quadraticI} \end{equation}
in contrast with the results found in reference \cite{gavil}.

An important quantity that gives information about the behavior of the
fluctuations of the interface is the vector normal to the interface:
\begin{equation}
 n^{\mu}=\frac{(-a^2(t)f_{t},f_{x},f_{y},1)}
 {a(t)\sqrt{1+f_{x}^{2}+f_{y}^{2}-a^{2}(t)f_{t}^{2}}} \nonumber\\
\end{equation}
Its correlation functions, to be computed below, will give information on
whether departures from a flat interface are significant.

At this point we would like to compare our result with those of Garriga and
Vilenkin.  These authors found that the fluctuations perpendicular to the
interface are associated with instabilities that manifest themselves as a
tachyonic mass for these fluctuations.  From the discussion above, we are led
to conjecture that the appearance of this tachyonic mass is the result of a
quantization that does {\it not\/} preserve translational invariance.

Because our procedure {\it does\/} preserve translational invariance but is
non-covariant in the intermediate steps, a direct comparison of our results is
somewhat subtle.  However, we can gain some insight by trying different
parametrizations of our kink-profile.  Consider the following the
parametrization
\begin{eqnarray}
 \eta(z,f(x,y,t),t) & = &  \eta_s(\chi(z,x,y,t)) \label{etachi2} \\
 \chi (z,x,y,t)      & = &
 \frac{a(t)z-F(x,y,t)}{\sqrt{1+a^{-2}(F_{x}^{2}+F_{y}^{2})-F_{t}^{2}}}
\label{chibreak} \end{eqnarray}
instead of that of equations (\ref{etachi},\ref{goch}).

We see that $\chi$ is {\it not\/} invariant under the the rigid translation
$F\rightarrow F+F_0,\,z\rightarrow z+F_0$; instead a translation of the
interface is compensated by a {\it time dependent\/} transformation $F
\rightarrow F+F_0/a(t),\,z\rightarrow F+f_0$.  Because of the time derivative
terms in the original action such a transformation is not an invariance of the
action, which is then changed by terms proportional to $h$ (time derivatives of
the scale factor), thus breaking translational invariance.  Following the same
steps leading to the effective action found above, neglecting higher derivative
terms, integrating by parts and rearranging terms we arrive at:
\begin{eqnarray}
 I&=&-\frac{1}{\lambda}\int {\rm d}^3 x\; a^2(t)(C_0-C_2 h^2/2)\sqrt{1+a^{-2}
  (F_{x}^{2}+F_{y}^{2})-F_{t}^{2}} \nonumber\\
 &+&\frac{C_0}{2\lambda} \int {\rm d}^3 x\;  F^2\,\frac{a^2 h^2
  \left(3+\dot{h}/h^2\right)}
  {\sqrt{1+a^{-2}(F_{x}^{2}+F_{y}^{2})-F_{t}^{2}}}+{\cal{O}}(F^2F_iF_{jk}).
\end{eqnarray}

Keeping only the quadratic terms in the action we find that the fluctuation
field acquires a tachyonic mass, $m_F^2=-3h^2$.  This is the same value of the
mass obtained by Garriga and Vilenkin \cite{gavil}.  Thus we conjecture that
the scalar field that measures departures from a flat interface introduced by
Garriga and Vilenkin is equivalent (at least to lowest order in derivatives) to
the scalar field $F$ parametrizing the fluctuations of the interface as in
equations (\ref{etachi2},\ref{chibreak}).  As explained above, this
parametrization explicitly breaks rigid translational invariance in any FRW
cosmology.  The appearance of the mass term is understood as a consequence of
this explicit breakdown of translational invariance, although this fact does
not explain the tachyonic nature of the mass.

Thus it seems to us that there are advantages and disadvantages in both
formulations.  Whereas the formulation of Garriga and Vilenkin is desirable in
that it maintains explicit covariance, there is the feature of instabilities
associated with the tachyonic mass of the fluctuations, which if our analysis
is correct, indicates the breakdown of translational invariance in the
quantization procedure.  On the other hand, our formulation, in terms of
collective coordinate quantization, sacrifices explicit covariance, although
the final result is covariant, but explicitly treats translational invariance
and its restoration via the collective coordinate quantization. The collective
coordinates represent massless fields as a consequence of this translational
invariance.

\subsection{Quantization of the Fluctuations}

Quantizing the fluctuations of the interface allows us to answer some relevant
questions about the dynamics of the interface.  In particular we can
answer the question that we posed at the beginning of the section that is
whether the interface (wall) is flat or strongly fluctuating
at long distances. In order to answer this question we must compute the
correlation function of the vectors normal to the interface at long distances
$1 \ll r \leq h^{-1}$ with $r$ the distance on the two-dimensional surface of
the interface.

Since we are interested in long distance physics, we will only consider slowly
varying fluctuations of the fluctuation field $f$ and neglect higher derivative
terms in the action, keeping only the quadratic terms (one can be brave and
pursue a perturbative expansion but we will content ourselves here with a
lowest order calculation) in the action.  Repeating eq.~(\ref{quadraticI}),
\begin{equation}
 I \simeq -\frac{1}{2\lambda}\int {\rm d}^3 x\; a^{2}(t)
 \left[C_{0}-C_{2}(5+2\dot{h}
 /h^{2})h^{2}/2\right]\left[f_{x}^{2}+f_{y}^{2}-a^{2}(t)f_{t}^{2}\right]
\end{equation}
The equations of motion for the field $f$ are given by:
\begin{equation}
 f_{tt}+f_{t}\left\{4h-\frac{C_2(5h\dot{h}+\ddot{h})}{\left[C_0-C_{2}(5+2
 \dot{h}/h^{2})h^{2}/2\right]}\right\}-\frac{\Delta f}{a^2(t)}=0
\label{eqofmotion} \end{equation}
with $\Delta$ being the two-dimensional Laplacian.

In the most general case, the time dependence of the above equation is far too
complicated to pursue analytically and one would have to resort to numerical
integrations.  Thus we concentrate on the case of de Sitter expansion with
scale factor $a(t)=e^{ht}$ that allows an analytic treatment.

The fluctuation field is expanded in terms of creation and annihilation
operators and the mode functions which are solutions of the above equation of
motion:
\begin{eqnarray}
 f(\vec{x},t)& = & \frac{1}{\sqrt{\sigma A}}\sum_p \left[a_p
 e^{i\vec{p}\vec{x}}v_p(t)+
 a_p^{\dagger} e^{-i\vec{p}\vec{x}}\,v_p^*(t)\right] \label{flucexp} \\
 \sigma  & = & \frac{1}{\lambda}\left[C_{0}-5C_{2}\frac{h^2}{2}\right]
\label{surftens} \end{eqnarray}
where the ``surface tension'' $\sigma$ has been absorbed in the definition of
$f$ to make it canonical and $A$ is the (comoving) area of the (planar)
wall.

Performing the change of variables on the mode functions
$v_p=e^{-2ht}\chi_p(t)$ the equation for
$\chi_p$ reads:
\begin{equation}
 \frac{\partial^2\chi_p}{\partial t^2}+
 \left(\frac{\vec{p}^{\,2}}{a^2(t)}-4h^2\right)\chi_p=0.
\end{equation}
whose solutions are linear combinations of the Bessel functions.  The mode
functions $v_p$ are
\begin{equation}
 v_p(t)= \sqrt{\frac{\pi}{4h}}
e^{-2ht}\left[A_p\, H^{(1)}_{2}\left(\frac{p e^{-ht}}{h}\right)+
 B_p\, H^{(2)}_{2}\left(\frac{p e^{-ht}}{h}\right)\right].
\end{equation}
where the coefficients $A_p,\  B_p$ are arbitrary so far.

Imposing canonical commutation relations between $f$ and its canonical
conjugate momentum $\Pi_{f}={\delta {\cal{L}}}\slash{\delta f_{t}}=\sigma
f_{t}a^4 $ leads to the relation
\begin{equation}
 |A_p^2|-|B_p^2|=1.
\end{equation}

It is a well known feature of quantization in curved spacetimes \cite{birrel}
that a choice of $A_p$ corresponds to a choice of vacuum state.  Although we
do not have a physical criterion to pick a particular vacuum state, we will
choose the Bunch-Davies \cite{bunch} vacuum for simplicity.  Such a choice
implies
\begin{equation}
 B_p=0. \nonumber
\end{equation}

Without loss of generality we can take $A_p=1$.  Finally, the field $f$ is
expanded in creation and annihilation operators with respect to the
Bunch-Davies vacuum state as:
\begin{eqnarray}
 f(x,y,t)= e^{-2ht}\sqrt{\frac{\pi}{4h\sigma A}}\sum_p
 \left[a_pe^{i\vec{p}_\perp\cdot
 \vec{x}_\perp}H^{(1)}_{2}\left(\frac{pe^{-ht}}{h}\right)+
 a_p^{\dagger}e^{-i\vec{p}_\perp\cdot\vec{x}_\perp}
 H^{(2)}_{2}\left(\frac{pe^{-ht}}{h}\right)\right]
\end{eqnarray}

As mentioned above, we are interested on the long-wavelength fluctuations of
the interface at long distances and at long times after its formation.  Thus we
will study the regime $t \gg h^{-1} \; ; \; hr \leq 1$.  Furthermore there are
physical cutoffs that we must introduce: the (comoving) wavelengths cannot be
bigger than the horizon, and because the nature of our approximation cannot be
shorter than the correlation length (the expansion is in derivatives, thus
valid for slowly varying fields on the scale of the correlation length).
Therefore integrals over wavevectors will be restricted to the interval $h \leq
p \leq 1$.

The fluctuation field $f(x,y,t)$ measures the departure from a flat interface.
Thus a quantity of interest is the correlation function of the vector normal to
the space-like interface.  This normal vector is obtained from the induced
metric {\it on the two-dimensional surface\/} and given by
\begin{equation}
 \vec{n}=\frac{(-f_{x},-f_{y},1)}{\sqrt{1+f_{x}^2+f_{y}^2}}\simeq(-f_{x},
 -f_{y},1-f_{x}^2/2-f_{y}^2/2). \label{normal}
\end{equation}
The equal-time two-point correlation function is given by
\begin{equation}
 <\vec{n}(\vec{x},t)\cdot \vec{n}(\vec{y},t)>
 \simeq 1+\frac{\pi^2e^{-4ht}}{2h\sigma}
 \int_{h}^{1} dp p^3 [J_{0}(pr)-1]\left|H^{(1)}_{2}
 \left(\frac{pe^{-ht}}{h}\right)\right|^2.
\end{equation}
notice from  eq. (\ref{normal}) that
$<\vec{n}(\vec{x},t)\cdot \vec{n}(\vec{x},t)> =1 $ (to the order considered)
and that the correlation function  does not require any  short distance
subtraction.  Its long time, long distance behavior is found to be
\begin{equation}
 <\vec{n}(\vec{x},t)\cdot\vec{n}(\vec{y},t)> \approx
 1-\frac{8\pi^2h^3}{\sigma}\ln r/2+\cdots
\label{normalcorr} \end{equation}
where $\cdots$ stand for sub-leading terms that fall off fast at large $r$.

It is instructive to compare this result with that in Minkowski spacetime, for
which $< \vec{n}\cdot\vec{n}> \approx 1-r^{-3}$.  Clearly the interface is
``rougher'' in de Sitter spacetime.  The result (\ref{normalcorr}) raises the
very interesting possibility of anomalous exponents in the correlation
function. Notice that for large $r$ the logarithmic (infrared) singularities
become strong and eventually would have to be resummed in order to obtain
meaningful long distance behavior.  {\it Assuming\/} such a resummation of the
lowest order result we obtain the long distance behavior:
\begin{eqnarray}
 <\vec{n}(\vec{x},t)\cdot\vec{n}(\vec{y},t)> & \approx &
 r^{-\alpha} \label{anoexp}\\
 \alpha & = &
 \frac{8\pi^2 h^3}{\sigma}.
\end{eqnarray}

This result has interesting implications.  In particular, we find
that at long times and distances the vectors normal to the interface are
uncorrelated and the interface is ``rough'' rather
than approximately flat with small fluctuations that fall-off rapidly at
long distances.

This situation is very similar to that of the X-Y model and other models in
statistical mechanics in $1+1$ Euclidean dimensions (typically massless scalar
field theories) in which logarithmic singularities sum up to anomalous
dimensions \cite{polyakov} much in the same way.  This would indeed be a
tantalizing possibility that needs to be studied further, perhaps by keeping
higher order terms in the effective action and resumming using renormalization
group arguments. This is certainly beyond the scope of this article.

Another quantity of interest that measures properties of the
interface is the distortion \cite{gavil}:
\begin{equation}
 D^2(\vec{x},\vec{y};t) = <\left[f(\vec{x},t)-f(\vec{y},t)\right]^2>=
 2<f^2(\vec{x},t)>-2<f(\vec{x},t)f(\vec{y},t)>
\end{equation}
We evaluated this correlation function for large times and distances ($t \gg
1/h \;;\; r \gg 1 \gg hr$) and found
\begin{eqnarray}
 D^2(\vec{X},\vec{Y})& \approx & \frac{4\pi^2h^3}{\sigma} r^2\ln r.
\end{eqnarray}
This result is consistent with that of the normal-normal correlation function
(\ref{normalcorr}).

This correlation function shows once again that the interface is strongly
fluctuating at long distances and cannot be considered flat.

\section{Conclusions}

In this article we have focused on the properties of domain walls in general
FRW Universes.  These defects will appear during phase transitions in typical
inflaton theories and ultimately drive the dynamics of the process of phase
separation. They may also contribute to density fluctuations and perhaps to
structure formation.

Our results present a rich picture of the properties of these defects.  After
analyzing numerically the case of de Sitter spacetime and assuming that the
boundary conditions on the field configuration require that the horizon size be
larger than the typical correlation length we set up a systematic perturbative
expansion in powers of $h=H/m$ and worked out in detail the first order
correction from curved spacetime effects.  We see then that even in situations
in which the Universe is expanding slowly enough to allow us to set up our
perturbation theory for defects, new phenomena occurs with definite
cosmological implications.

In particular, we argue that in a general FRW spacetime, consistency
(renormalizability) of the theory requires a coupling to the Ricci scalar. With
this coupling the energy momentum tensor of defects can show some remarkable
new features, including a negative energy density at the origin. Clearly this
observation brings interesting possibilities for topological inflation near the
center of the defect that must be studied within the full set of Einstein's
equations in the presence of this configuration. For some values of the
coupling to the Ricci scalar, we find that the energy density at the center is
{\em lower} than at a distance of the order of the correlation length away from
the center.

Such a behavior of the energy momentum tensor may provide an interesting
mechanism for scalar density perturbations.

Using collective coordinate quantization of the fluctuations perpendicular to
the wall, we obtained the effective Lagrangian for these fluctuations in the
long-wavelength approximation. There are several noteworthy features of this
effective Lagrangian: curved spacetime effects renormalize the surface
tension, and we find systematic corrections to the ``Nambu'' action.

We have also shown how to quantize the fluctuations about a completely flat
wall, and how the collective coordinate procedure yields the expected Goldstone
mode associated with translational invariance.  This analysis then allows us to
compute correlation functions and to see that, at least in the de Sitter case,
fluctuations are fairly strong at long distances and long times after formation
of the defect leading to conclusion that the surface tends to be rougher than
in flat spacetime. These results and the method developed for the quantization
of the fluctuations may yield some insight on novel mechanisms to study
density perturbations.

What we have done for walls can also be done for other defects, such as cosmic
strings.  These should exhibit some interesting behavior, especially for the
fluctuations, since there will now be collective coordinates associated with
the internal $U(1)$ symmetry whose spontaneous breaking gives rise to the
string.  Interesting possibilities are corrections to the string tension from
curved spacetime effects and also corrections to the ``Nambu-Goto'' action
{}from higher derivative terms, and perhaps novel behavior of the energy
momentum tensor at the core of the defect when the coupling to the Ricci scalar
is introduced.  We are currently studying these and other issues.

\newpage
\leftline{\Large\bf Acknowledgments} \bigskip
DB was partially supported by NSF Grant \# PHY-9302534 and NSF Grant
INT-9016254 bi-national collaboration with Brazil.  FT was partially supported
by CNPq.  DEB and RH were partially supported by the U.S. Dept.\ of Energy
under Contract DE-FG02-91-ER40682.  AGR thanks the Spanish M.E.C. for support
under an FPI grant.  FT and AGR thank the Dept. of Physics, Univ. of Pittsburgh
for hospitality.  DB and RH would like to thank E. Kolb for inspiring the
problem and for fruitful conversations and remarks. DB would like to thank
Professor A. Vilenkin for illuminating conversations and comments.
\bigskip



\newpage
\INSERTFIG{3.95}{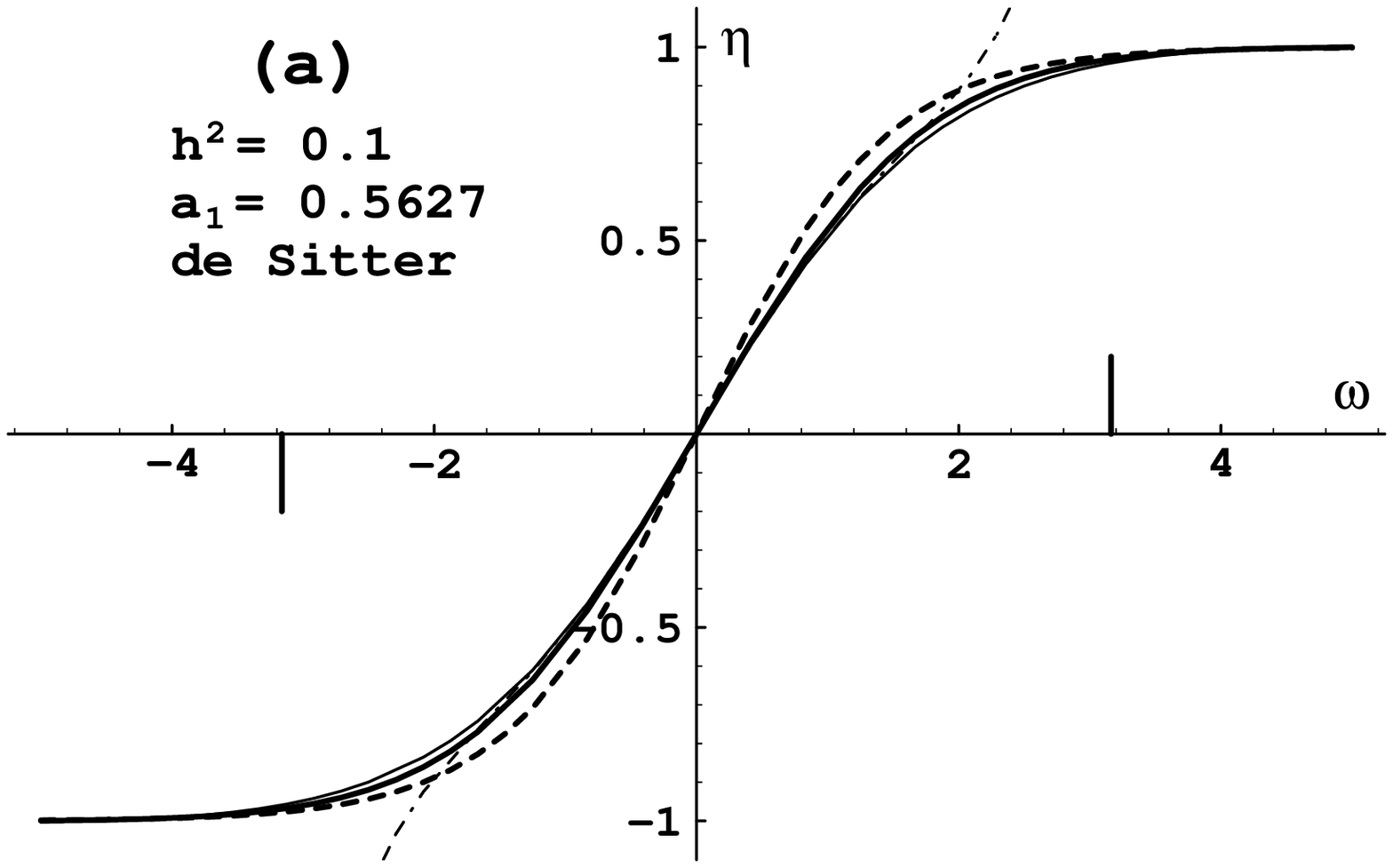}{~}
\INSERTFIG{3.95}{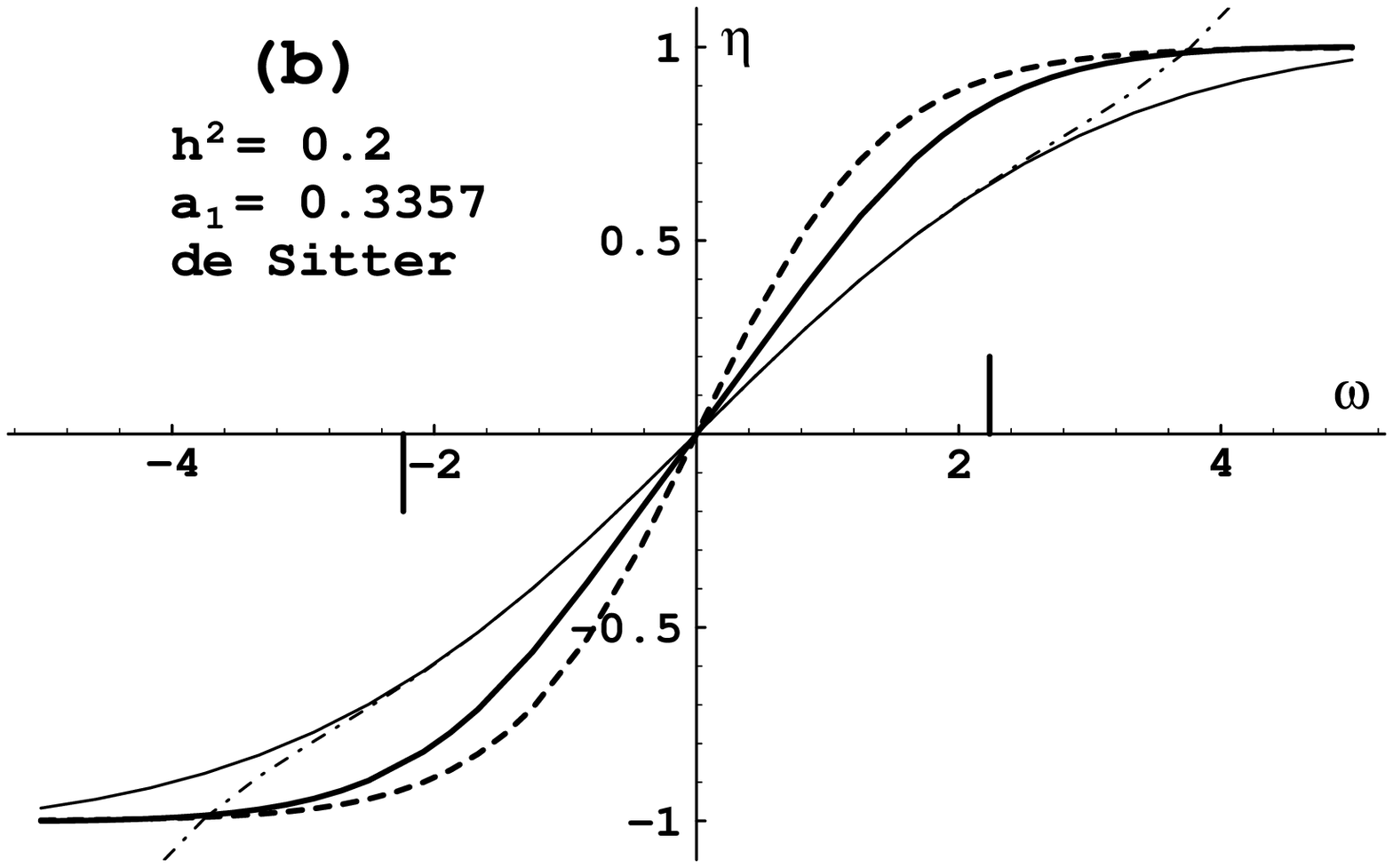}{~}
\INSERTFIG{3.95}{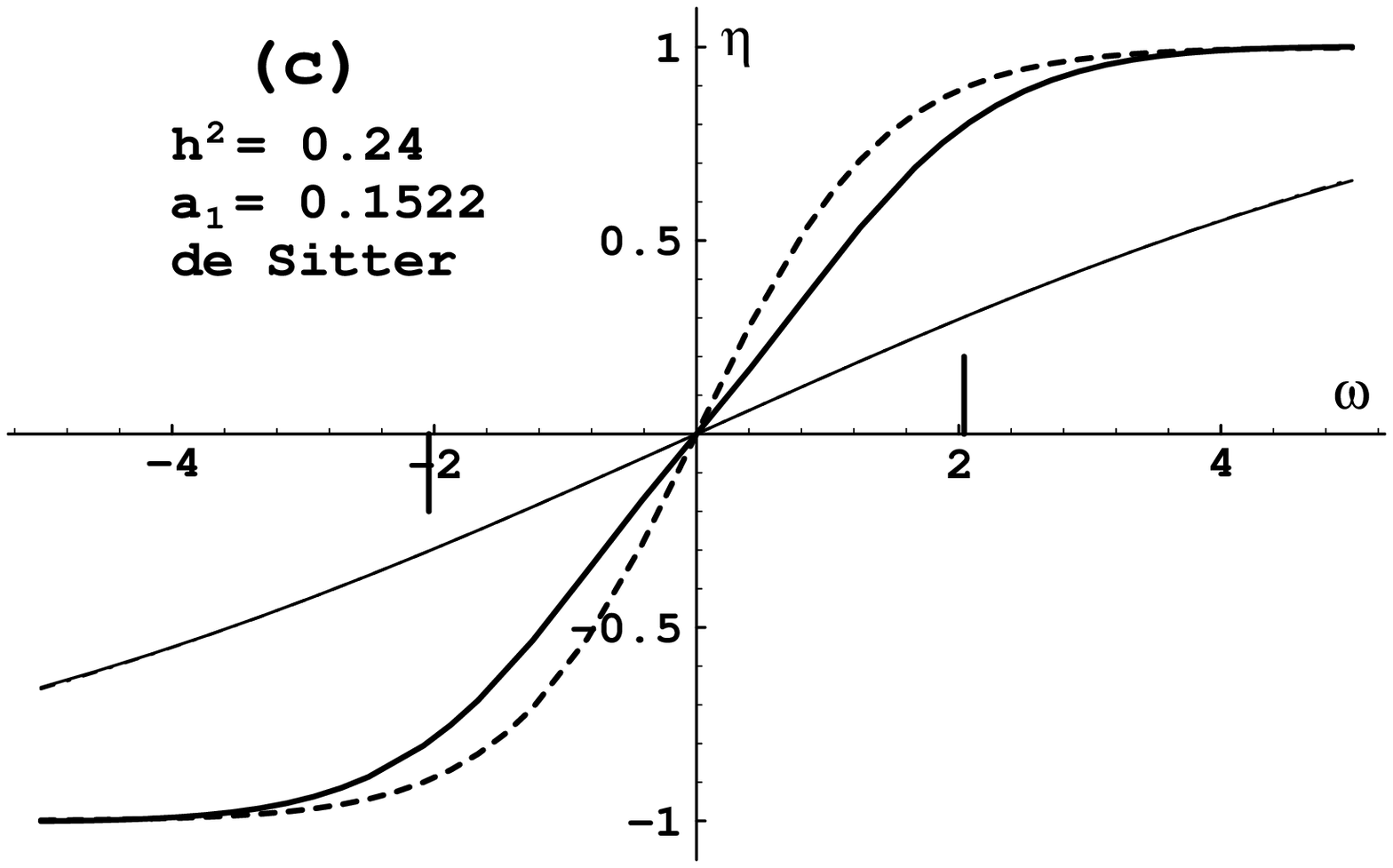}{Fig.~1: Domain wall profiles for de Sitter space,
  $h^2$ = \{.1, .2, .24\}.  Dash=kink, thin=exact, thick=small-$h^2$,
  dot-dash=small-$\omega$.  $\omega=\pm 1/h$ is marked by vertical ticks.}
\vskip.3in
\INSERTFIG{5.}{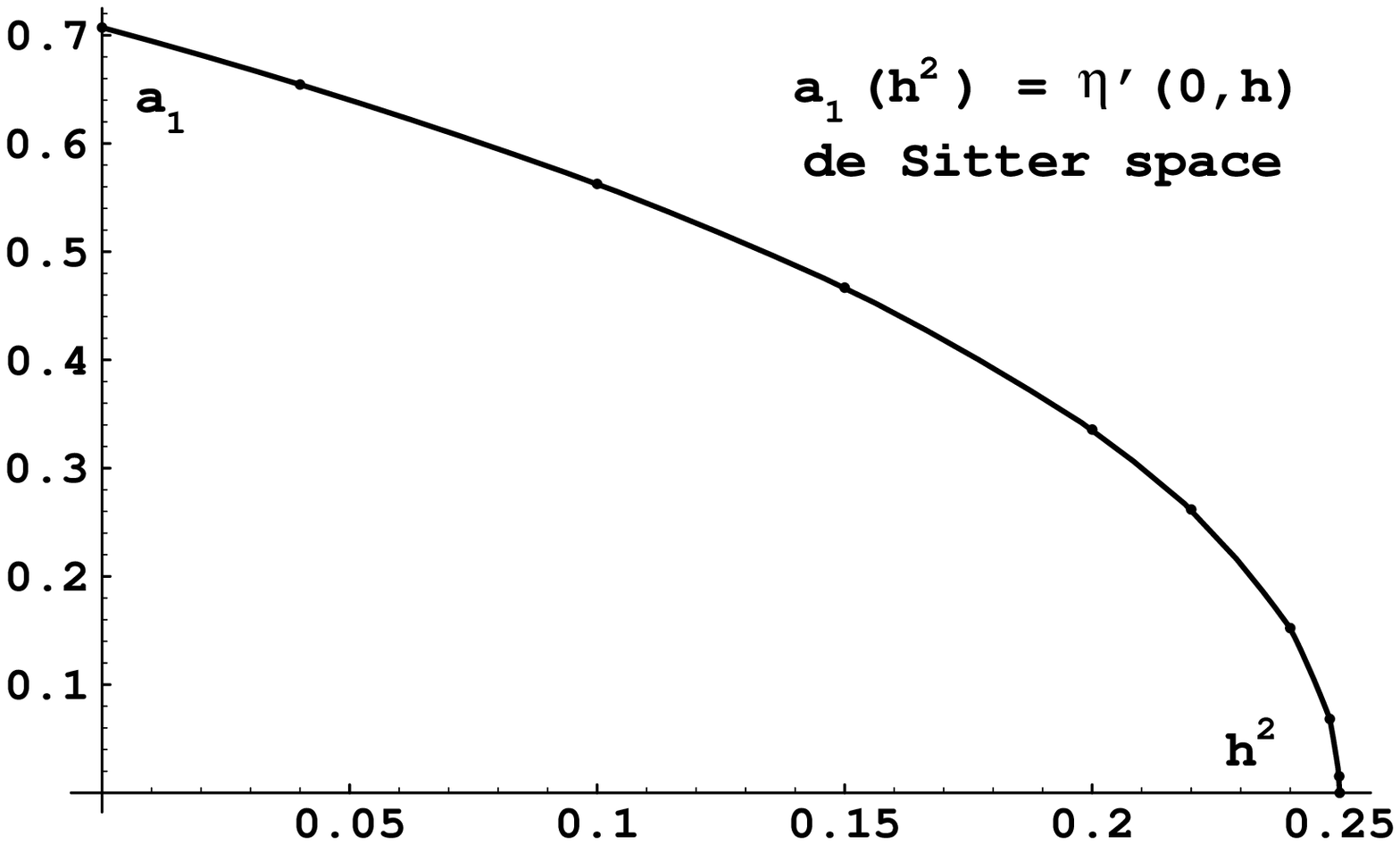}{Fig.~2: $a_1 \equiv \eta'(0,h)$ vs. $h^2$ for de
  Sitter space.}
\vskip1in
\INSERTFIG{5.}{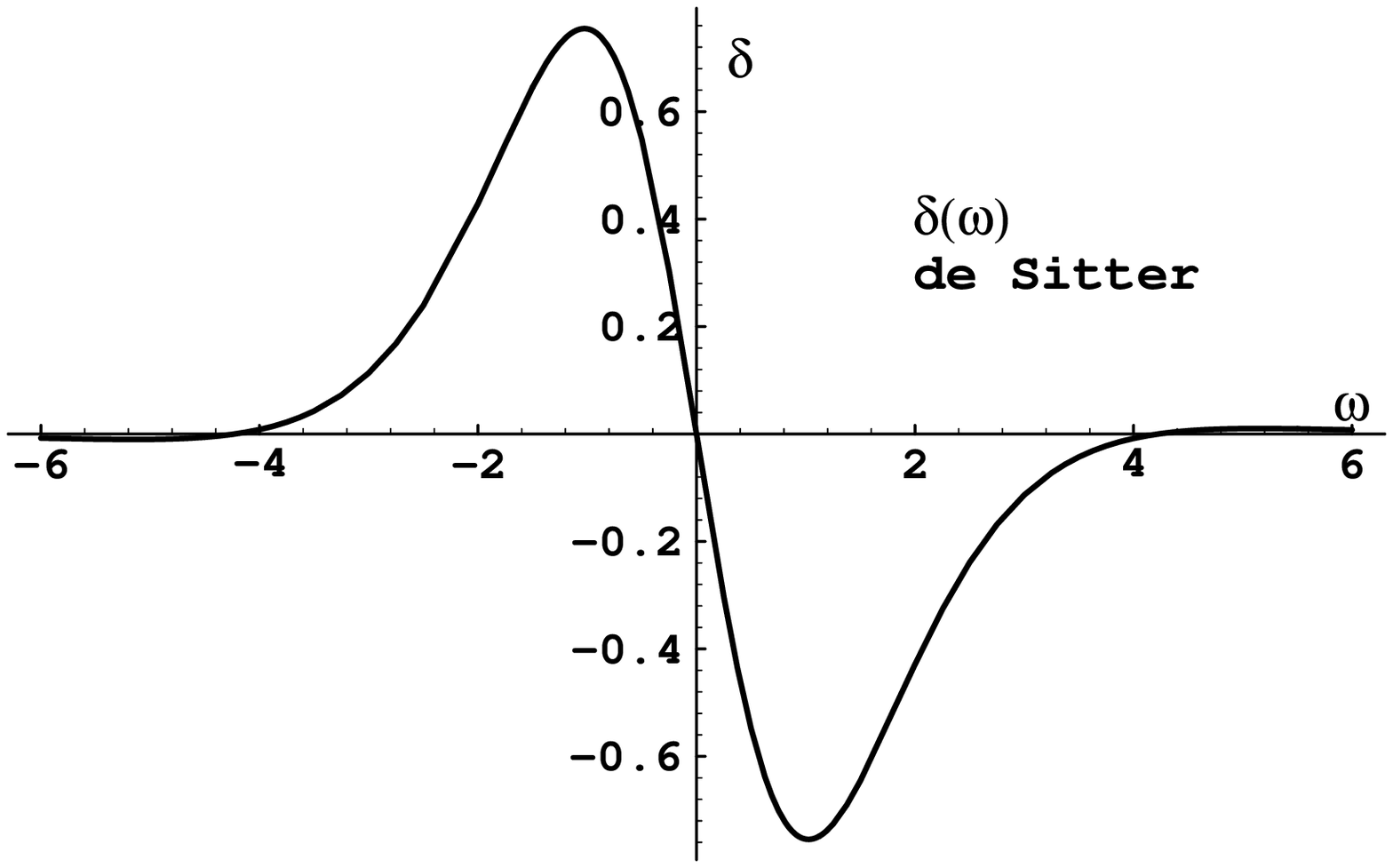}{Fig.~3: $\delta(\omega)$ for de Sitter space.}
\INSERTFIG{5.}{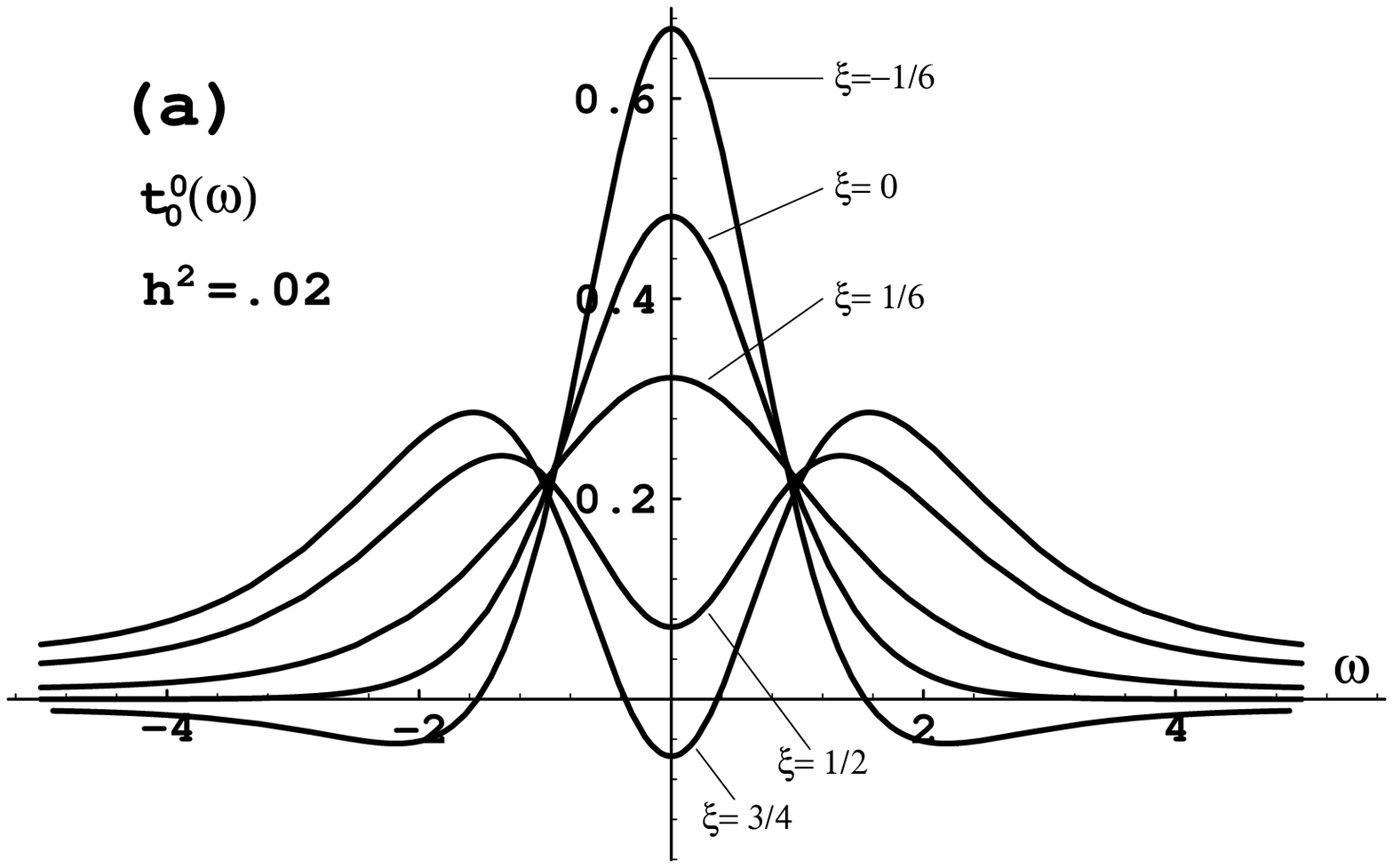}{~}
\INSERTFIG{5.}{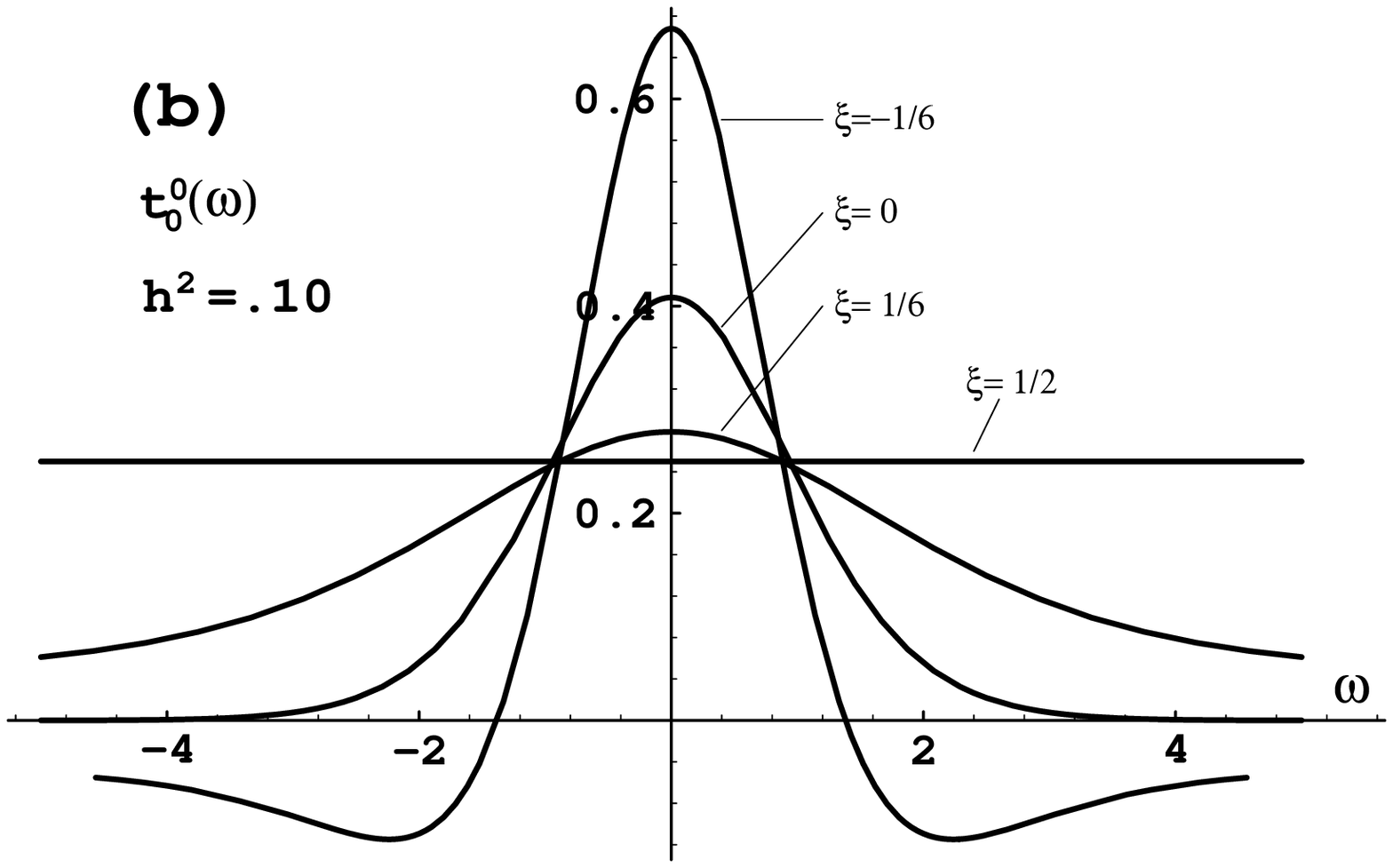}{Fig.~4: $t_0^{\;0}(\omega)$ in de Sitter space,
  various $\xi$, $h^2$ = \{.02, .1\}.}

\end{document}